\documentclass[aps, twocolumn]{revtex4-1}
\usepackage{amsfonts}
\usepackage{amsmath}
\usepackage{color} 
\usepackage{amssymb}
\usepackage{graphicx}
\usepackage{times}
\usepackage{physics}
\usepackage[colorlinks=true,linkcolor=blue,citecolor=red,plainpages=false,pdfpagelabels]
{hyperref}
\usepackage[encapsulated]{CJK}
\usepackage{color}

\begin{document}

\title{Neural Network Enhanced Single-Photon Fock State Tomography}
\author{Hsien-Yi Hsieh,$^{1}$, Yi-Ru Chen,$^{1}$, Jingyu Ning,$^{1}$ Hsun-Chung Wu,$^{1}$ Hua Li  Chen,$^{2}$ Zi-Hao Shi,$^{1}$ Po-Han Wang,$^{3}$ Ole Steuernagel,$^{1}$ Chien-Ming Wu,$^{1}$ and  Ray-Kuang Lee$^{1,2,3,4,5}$}
\affiliation{$^{1}$Institute of Photonics Technologies, National Tsing Hua University, Hsinchu 30013, Taiwan\\
$^{2}$Department of Physics, National Tsing Hua University, Hsinchu 30013, Taiwan\\
$^{3}$Department of Electrical Engineering, National Tsing Hua University, Hsinchu 30013, Taiwan\\
$^{4}$Center for Theory and Computation, National Tsing Hua University, Hsinchu 30013, Taiwan\\
$^{5}$Center for Quantum Science and Technology, Hsinchu 30013, Taiwan}
 \email{rklee@ee.nthu.edu.tw}

\date{\today}
\begin{abstract}
Even though heralded single-photon sources have been generated routinely  through the spontaneous parametric down conversion, vacuum and multiple photon states  are unavoidably involved. 
With machine-learning, we report the experimental implementation of single-photon quantum state tomography by directly estimating  target parameters.
Compared to the Hanbury Brown and Twiss (HBT) measurements only with clicked events recorded, our neural network enhanced quantum state tomography characterizes the photon number distribution for all possible photon number states from the balanced homodyne detectors.
By using the histogram-based architecture, a direct parameter estimation on the negativity in Wigner's quasi-probability phase space is demonstrated.
Such a fast, robust, and precise quantum state tomography provides us a crucial diagnostic toolbox for the applications with single-photon Fock states and other non-Gaussisan quantum states.
\end{abstract}

\maketitle

\section{Introduction}
Quantum state tomography (QST) refers to the methodology in reconstructing the unknown quantum state with the acquired experimental data~\cite{QST-book, RMP-CV}.
The  maximum likelihood estimation (MLE) for QST finds the best-fitted probability distribution by treating the whole density matrix as the target of estimation~\cite{Banaszek, Hradil, Lvovsky}. 
As long as a sufficient computational effort is applied, MLE consistently yields a robust estimation, with the effectiveness in estimation strongly depending on the quantity of available data. 
Nowadays, QST has been successfully implemented as a  diagnostic toolbox both for many qubits (or qudits) systems in higher dimensions and for continuous variables in infinite dimensions~\cite{QST-Furusawa, QST-atom1, QST-atom2, QST-ion-1, QST-ion-2, QST-SC}.

However,  MLE suffers from the overestimation problem as the required amount of measurements to reconstruct the quantum state  exponentially  increases with the number of involved modes.
To overcome the overestimation problem  in MLE, several modified algorithms are proposed,  such as permutationally invariant tomography~\cite{permu}, quantum compressed sensing~\cite{compress}, tensor networks~\cite{tensor-1, tensor-2}, generative models~\cite{generative}, and~restricted Boltzmann machine~\cite{RBM}, by~assuming some physical restrictions imposed upon the state in question.
Moreover, unavoidable coupling from the noisy environment makes the reconstructions on the density matrix  with state degradation embedded, resulting in dealing with a non-sparse matrix  in a larger Hilbert space.

With the power to find the best fit to arbitrarily complicated solutions,   machine-learning (ML) enhanced QST has demonstrated its advantages in extracting complete information about the quantum states~\cite{RBM, QML-1, QML-2, PRL-22, GAN}.
Furthure more, instead of using the reconstruction model in training a truncated density matrix, with ML one may directly  generate the target parameters  with a  supervised characteristic model~\cite{Symmetry}.
Such a characteristic model-based ML-QST can be easily installed on edge devices such as FPGA, serving as an in-line diagnostic toolbox for all the possible applications.
As an example, this ML-QST has also been applied to the reconstruction of Wigner current~\cite{current}, demonstrating  experimentally  quantum dynamics in phase space  in great detail.
Compared to the time-consuming MLE, ML-QST paves the road toward a real-time and online QST~\cite{current, online}.

With the benefits from the good properties of the Gaussian states, including vacuum and squeezed states,   a neural network can directly analyze the {\it raw} data to obtain  the first and second moments of probability density function.
By applying the well-developed methods in  pattern recognitions~\cite{generalizability, Pattern-1, Pattern-2, Pattern-3}, one can easily deal with various Gaussian states,  producing a single scan QST in  speeding up data acquisition and data processing~\cite{PRL-22}. 
Nevertheless, difficulties arise for such a  relatively simple prediction map when non-Gaussian states are attacked. 
One may increase the number of neurons in dealing with non-Gaussian states, however the training process tends to cause overfitting problem. 

In the family of non-Gaussian states, single-photon Fock states play the core role as  photonic qubits to carry quantum information encoded~\cite{single-photon}.
Although the request for an on-demand source of single photons has led to intense research into developing truly deterministic single-photon states, {\it heralded} single-photon sources can be easily generated through correlated pairs of photons, by  detecting one photon (the heralding photon).
After the first experimental observation in 1970~\cite{SPDC}, nowadays, creating correlated photon pairs from spontaneous parametric downconversion (SPDC) has been routinely demonstrated with $\chi^{(2)}$ nonlinear crystals~\cite{typeII}.

To  characterize a single-photon Fock state, the common method is based on the second-order correlation function, $g^{(2)}(\tau)$, i.e.,  Hanbury Brown and Twiss (HBT) interferometer~\cite{HBT, Glauber, Sudarshan}.  
The  standard test for single-photon sources is a value of the second-order correlation function of the emitted field below $1/2$ at zero time delay, i.e., $g^{(2)}(0)<1/2$.
However, this criterion alone provides no information regarding the amplitude of the single-photon contribution for general quantum states. 
In particular, a low-intensity light source always has a vacuum contribution in the quantum state of light, cloaking actual single-photon projection~\cite{g2, low}.

In addition to HBT measurements, nonclassical effects in the single-photon Fock states can be demonstrated in phase space~\cite{sq-thermal-3},  such as a negative value in the Wigner function~\cite{9}.
Homodyne detection of the rotated quadratures provides an experimental implementation for the reconstruction on the Wigner function in  phase space, through the inverse Radon problem~\cite{10, 15}.
Tomographic  reconstruction of the single-photon states has been experimentally realized first  with phase-randomized pulsed optical homodyne tomography~\cite{single, instant, high-freq}, then with continuous temporal-mode matching~\cite{11}, toward having real-time and complete temporal characterization of a single photon~\cite{complete}. 
The development on homodyne tomography also provides a bridge between the single-photon and squeezed-vacuum states~\cite{bridge, antibunching, antibunching-loss}, as well as a methodology for various non-Gaussian states such as  two-photon Fock states and optical cat states~\cite{two-photon, cat}.

As the neural network predictor is often trained from some specific and limited amount of data, in this work, we develop the machine-learning enhanced single-photon Fock state tomography with  the histogram-based architecture.
 Histogram-based approaches are often used to reduce the computational cost~\cite{3}. 
 With an appropriately chosen bin-width for histogram, we demonstrate that the resulting quantum state reconstruction can still keep fidelity  high. 
 Further more, with the capability of hybrid quantum-classical neural networks or quantum neural network, the improvement in increase accuracy while reducing computational resources is also possible with quantum machine-learning~\cite{QML}.

The paper is organized as follows: in Section \ref{exp}, we introduce our experimental setup to perform the homodyne detections on heralded single-photon Fock states, generated from SPDC process inside a bow-tie cavity.
Then, the~implementations of the histogram-based neural network are illustrated in Section \ref{nn}. 
The comparisons on the predicted photon-number distributions from MLE and neural network, also as a function of the SPDC pumping power, are given in Section \ref{result}.
Moreover, a direct parameter prediction on the negativity is demonstrated, validating the feature extraction from our direct parameter estimations.
Finally, we summarize this work with some perspectives in Section~\ref{conclusion}.

\begin{figure}[t]
\includegraphics[width=8cm]{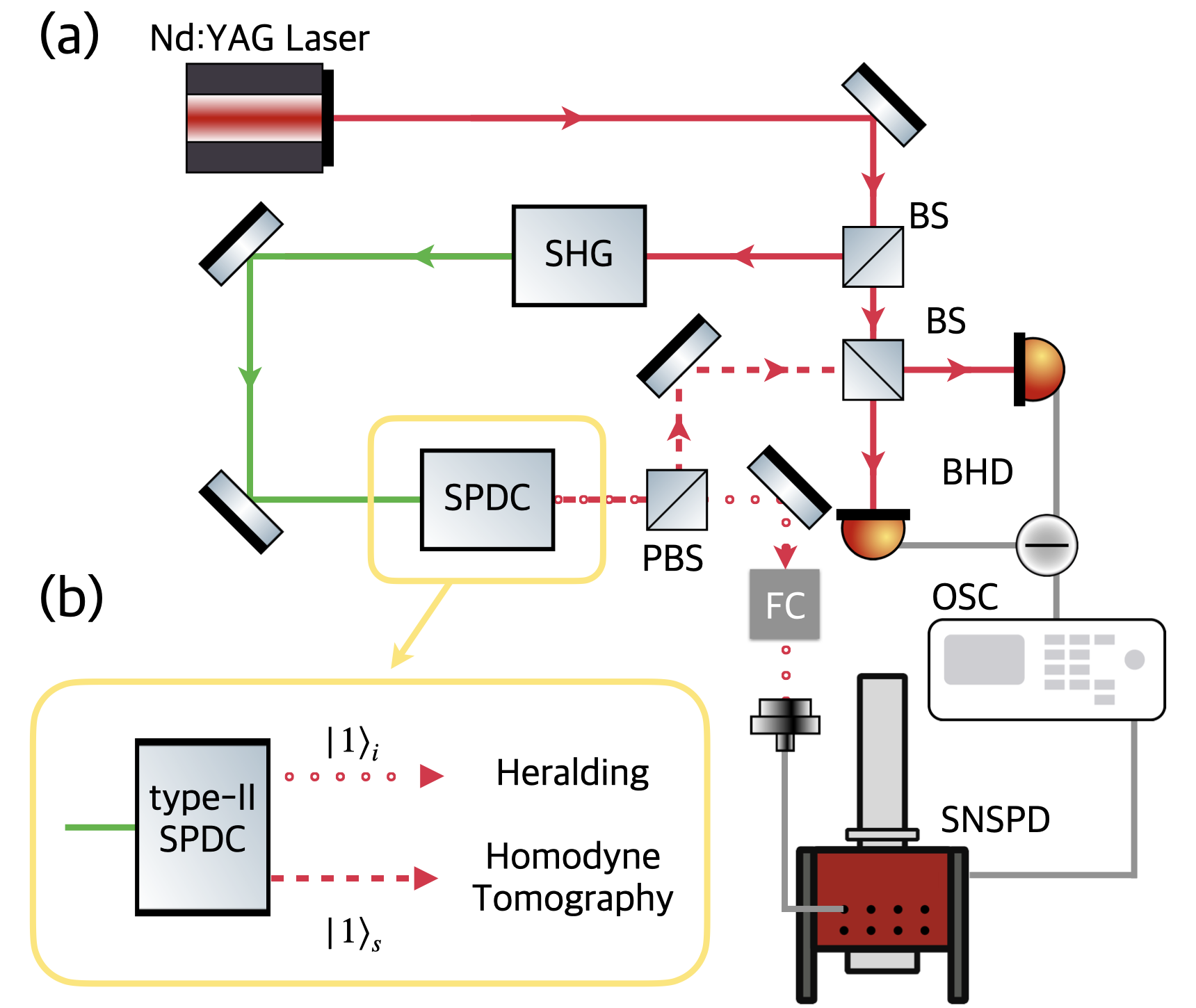}
\caption{(a) Our  experimental setup to generate single-photon Fock state and its quantum state tomography. (b) A simple schematic for generating heralded single photon precess by spontaneous parametric down conversion (SPDC) process. Here, SHG refers to the second harmonic generator; BHD: balanced homodyne detector; OSC: oscilloscope; SNSPD: superconducting nanowire single photon detector; FC: filter cavities; BS: beam splitter; PBS: polarizing beam splitter.}
\label{Fexp}
\end{figure}

\begin{figure}[t]
\includegraphics[width=8.0cm]{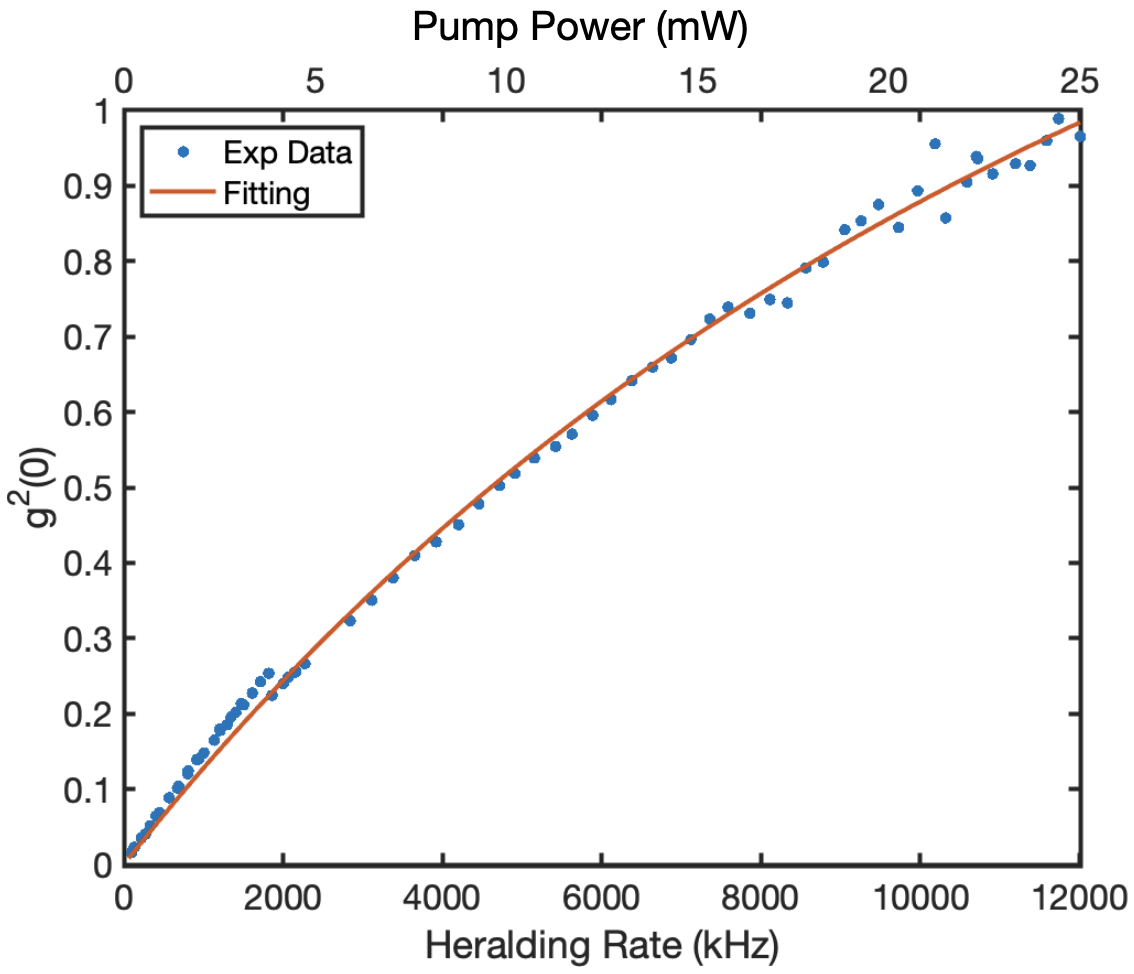}
\caption{Measured data on the second-order correlation function, $g^{(2)}(0)$, as a function of the recored heralding rate  in kHz,  with the corresponding pump power labeled (on top) in mW.}
\label{Fg2}
\end{figure}
\begin{figure*}
\includegraphics[width=16.0cm]{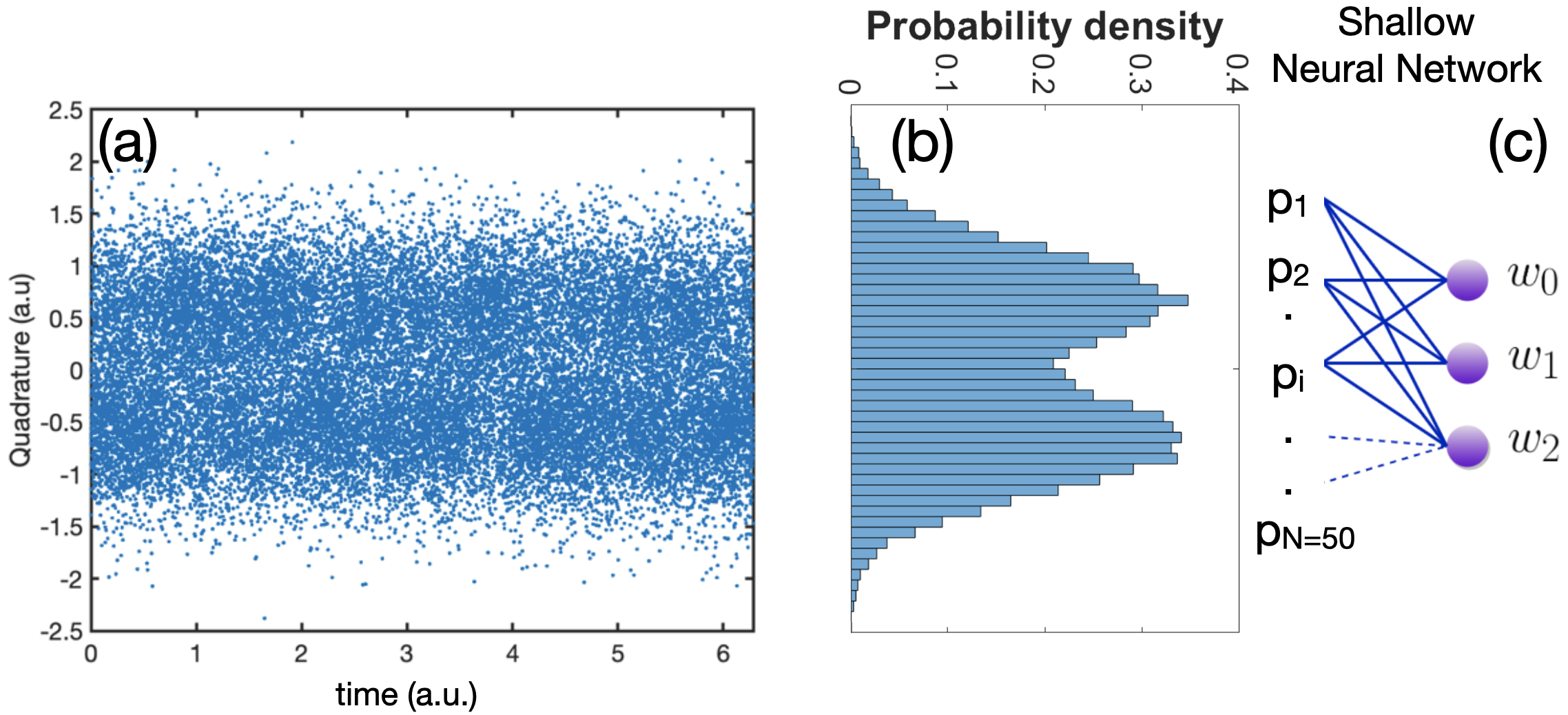}
\caption{(a) The time sequence of recorded  BHD raw quadrature  data measured from the oscilloscope. Here, the SPDC pump power is 3 mW. (b) The histogram of the corresponding probability density distribution, $p_i$ ($i = 1 \dots N$).
(c) With $N=50$ inputs, a shallow neural network is applied to generate directly  the predicted probability for different photon numbers $w_n$, with $n = 0, 1, 2$.}
\label{Fhisto}
\end{figure*}

\begin{figure}[t]
\includegraphics[width=8.0cm]{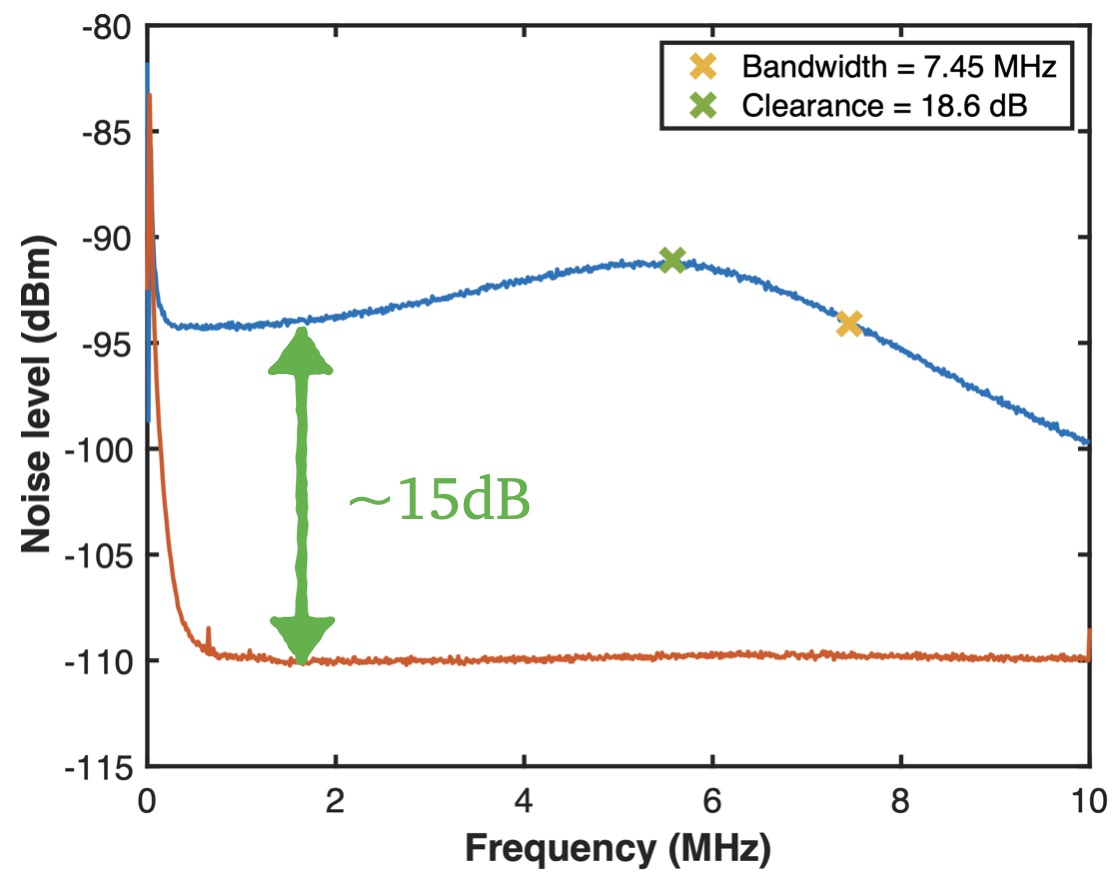}
\caption{Measured noise level in dBm of our balanced homodyne detector (BHD). Here, the spectrum for dark noise is depicted in Red color, up to 10 MHz. When the local oscillator is operated at 30 mW, the spectrum of noise level is depicted in Blue color, illustrating  a maximum clearance of 18.6 dB, along with a  3 dB bandwidth up to  $7.45$~MHz.}
\label{Fclear}
\end{figure}

\section{Experimental setup of our single photon quantum state tomography}
\label{exp}
The experimental setup for our heralded single-photon source and the quantum state tomography is illustrated in Fig.~\ref{Fexp}.
Here,  the main laser source is a continuous-wave (CW) Nd:YAG laser at the wavelength of 1064 nm. 
This laser is split  into two parts via a beam splitter (BS): one serves as the local oscillator (LO) beam for balanced homodyne detector (BHD); while the other one serves as the pump field for the second harmonic generator (SHG).
The SHG  provides the frequency doubling at 532 nm, through  a nonlinear crystal, i.e., periodically poled Lithium Niobate (PPLN), inside a bow-tie cavity.
Then, the green light of SHG signal is injected into another bow-tie cavity with a type-II PPLN crystal inside, in order to perform the spontaneous parametric down conversion (SPDC) process, which generates photon-pairs in two orthogonal polarizations. 
The full width at half maximum (FWHM) of our SPDC cavity is 31.8 MHz and the free spectral range (FSR) is 1.052 GHz. 
The output of orthogonal polarization beams, denoted as signal $|1\rangle_s$ and idler $|1\rangle_i$ photons in Fig.~\ref{Fexp}(b), are separated by a polarization beam splitter (PBS).

To ensure the mode-matching in degenerate modes, the idler photons from SPDC cavity is injected into a  filter cavity (FC) system with 6.5 MHz in bandwidth.
This FC system is composed of  a triangle cavity and two Fabry–Pérot cavities.
The detection on idler photons is performed by a superconducting nanowire single photon detector (SNSPD). 
Finally, the SNSPD  triggers the BHD to record the signal photons. 
As the Fock states are phase independent, we do not perform the measurement on rotated quadratures.
The extracted quadrature data are obtained by integrating the temporal mode function on the experimental data, i.e.,
\[
\hat X_i =  \int_\infty f(t) \hat x_i(t)\, dt,
\]
with $\hat x_i$ being  the $i$-th measurement data from the BHD.
Here,   the temporal mode function  $f(t)=\sqrt{\pi \gamma}e^{-\pi\gamma|t-t_c|}$ is described by the center time for a trigger event $t_c$, with the decay rate of SPDC, denoted as $\gamma$.

In Fig.~\ref{Fg2}, we report our measured data on the second-order correlation function at zero time delay, $g^{(2)}(0)$, as a function of the recored heralding rate from the detection on the heralding idler photon into SNSPD.
At the same time, the corresponding pump power (in mW) into the SPDC cavity is also depicted.
One can see clearly that our single-photon source demonstrates $g^{(2)} (0) < 1/2 $ when the pump power is smaller than 10 mW, or with the heralding rate about 4,000 kHz (4 MHz).

In addition to the  second-order correlation function, we also perform the quantum state tomography for the heralded single-photon state with  homodyne detection scheme.
In Fig.~\ref{Fhisto}(a), a typical time sequence of our noisy single-photon source is demonstrated from our experimental raw data measured from the oscilloscope.
Our BHD output is then integrated after convolution with a double-decayed (two time-constants) temporal mode function, along with a detection correction
treatment that assumes a detection efficiency of 92$\%$,  which is composed of the quantum efficiency of photodiodes ($99\%$), homodyne visibility efficiency ($96\%$), and the circuit noise of homodyne detection ($97\%$).

To make sure the measured noise level is not contaminated, our homodyne detectors are designed with a high common mode rejection ratio of more than $80$ dB~\cite{CMRR}.
As shown in Fig.~~\ref{Fclear}, when the local oscillator is operated at 30 mW the spectrum of noise level depicted in Blue color demonstrates a clearance of higher than 15 dB (with a maximum value of 18.6 dB).
However, our BHD only supports  a 3 dB bandwidth up to  $7.45$~MHz.

In the following, by considering the limit in our BHD bandwidth, we analyze the SPDC pump power up to 3 mW (or the heralding rate lower than 2 MHz) on our single-photon Fock state tomography with machine-learning. 
This operation condition also reflects the  scenario when the influence of vacuum significantly  enters into the actual single-photon projection, denoted as the low-intensity limit.

\section{Histogram-based machine-learning enhanced quantum state tomography}
\label{nn}

Before introducing the histogram-based machine-learning architecture, we conduct several tests on single-photon Fock state tomography by applying our previously developed convolution neural network for quantum state tomography with simulated {\it raw} quadrature data~\cite{PRL-22}. 
Here, more than 10,000 mock data set on noisy single-photon experiments are prepared, including different percentages of single-photon and vacuum states~\cite{11, single, high-freq, instant, complete}.
However, the inferred fidelity from these testing data set is limited to  $0.96$ even the input dimensions are up to  $10,000$.
One possible solution to improve the fidelity is to apply advanced machine learning architectures or to perform complex hyper parameter adjustments. 
 
As our goal is focusing on building a lightweight inference system which can be embedded into quantum optics experiments, working on {\it raw} quadrature data needs much more computation efforts in dealing with complicated data sets.
Instead, we  construct histograms to reduce the input data size, as well as the required computational cost. 

To reconstruct the quantum state in our  SPDC experiment, the corresponding tomographic data is the recorded event from our homodyne measurement, i.e., 
\begin{eqnarray}
p(X)=\sum_{n=0}^{\infty}w_{n}\frac{1}{\pi ^{1/2}2^{n}n!}H^{2}_{n}(X)e^{-X^{2}},
\label{eqpX}
\end{eqnarray}
with the Hermit polynomial $H_n (X)$.
Here, we already expand the probability  probability distribution in Fock (photon number) basis, with  $X$ being the value of rotated quadrature and $w_{n}$ being the photon number probability (weighting factor).
As the Fock states are independent to the quadrature phase, we also apply the phase-average measurement~\cite{10, 11,15, single, high-freq, instant, complete} to our homodyne data.
Tomographic reconstruction here is to estimate the photon number distributions $w_{n}$ from the measured quadrature data $X$.

By checking all the experimental data, we first set the quadrature value  between -3.2 and +3.2 as our range.  
Then, we divide this closed interval $[-3.2, 3.2]$ in the quadrature into 50 sub-intervals (bins), see Fig.~\ref{Fhisto}(b).
The relative frequency of the $i$-th bin is calculated by $f_{i}=\frac{N_{i}}{N}$, with $N_{i}$ denoting the counts in the $i$-th bin and $N$ being the total counts in the quadrature axis. 
We want to remark that in our numerical experiments $N = 50$ bins (sub-intervals) are enough to achieve good results.

The relative frequency of $i$-th bin, $f_i$, is  used to estimate the value of probability density defined on sampled quadrature values $\hat X_{i}$.
\begin{eqnarray}
\frac{f_i}{\Delta X}\simeq p_{i}(\hat X_{i}), \quad i=1 \dots N.
\label{eqfi}
\end{eqnarray}
Here, $\Delta X$ is the length of each bin and $\hat X_{i}$ is a specific point in $i$-th bin.
 With a uniform gird on $X$, the estimated value of probability density, $p_{i}$ is illustrated in Fig.~\ref{Fhisto}(b). 
 After this {\it binning} process, our tomographic problem now is transferred to  predict $w_{n}$ from the estimated value $p_{i}$ defined on the discrete grid $\hat X_{i}$. 
 When the number of input quadratures is large enough, the relative frequency converges to the probability, which enables a good approximated value of probability density $p_{i}$.
Otherwise, errors occur in the binning process.

In Fig.~\ref{Fhisto}(c), the schematic of this histogram-based neural network for single-photon Fock state QST is illustrated. 
Here, we apply a shallow neural network for 50 inputs from  the histogram-based  inputs, i.e., the estimated values of quadrature probability density $p_i$. 
Our learning task is to build a map supporting multiple instance setting:
\begin{eqnarray}
p_{i} \rightarrow w_{n}.
\label{eqpi}
\end{eqnarray}
Then, the outputs can generate directly  the predicted probability for different photon numbers $w_n$.

To  train our prediction map  inferring different quantum states from various tomographic data, we feed  machine with well prepared  training data set $\left\{p^{k}_{i},w^{k}_{n} \right\}$.
Here the index $k$ counts for different instances which describe specific quantum states. 
In this learning task, we use 10,000 training data set ($k= 10,000$) and  another 10,000 testing data with different weighting values of $w_{0}$, $w_{1}$, $w_{2}$, i.e., vacuum, single-photon, and two-photon Fock states, respectively.
A uniform distribution  in $[0, 1]$  is sampled. 
Further more,   by considering the low-intensity condition in our SPDC experiments, we let $w_{0}+w_{1}+w_{2}=1$ without other multi-photon events. 
With the simulated data as the ground truth, our histogram-based neural network can ensure the average fidelity higher than 0.999 with 10,000 instances in testing data sets.

In our single-photon Fock state QST, we also remark that the prediction map can be builded only with a shallow neural network, see Fig.~\ref{Fhisto}(c).
Additional hidden layers are not needed here. 
We also perform the numerical test confirming  that the neural network can maintain a good performance without any introduction of nonlinear activation functions.
In the training process, we train 10 epochs such that the mean squared loss of both training and testing data decreases to $10^{-7}$. 
The optimization process employs the {\it Adam optimizer} with default hyper-parameter settings, including a learning rate of 0.01.

\begin{figure}[t]
\includegraphics[width=8.4cm]{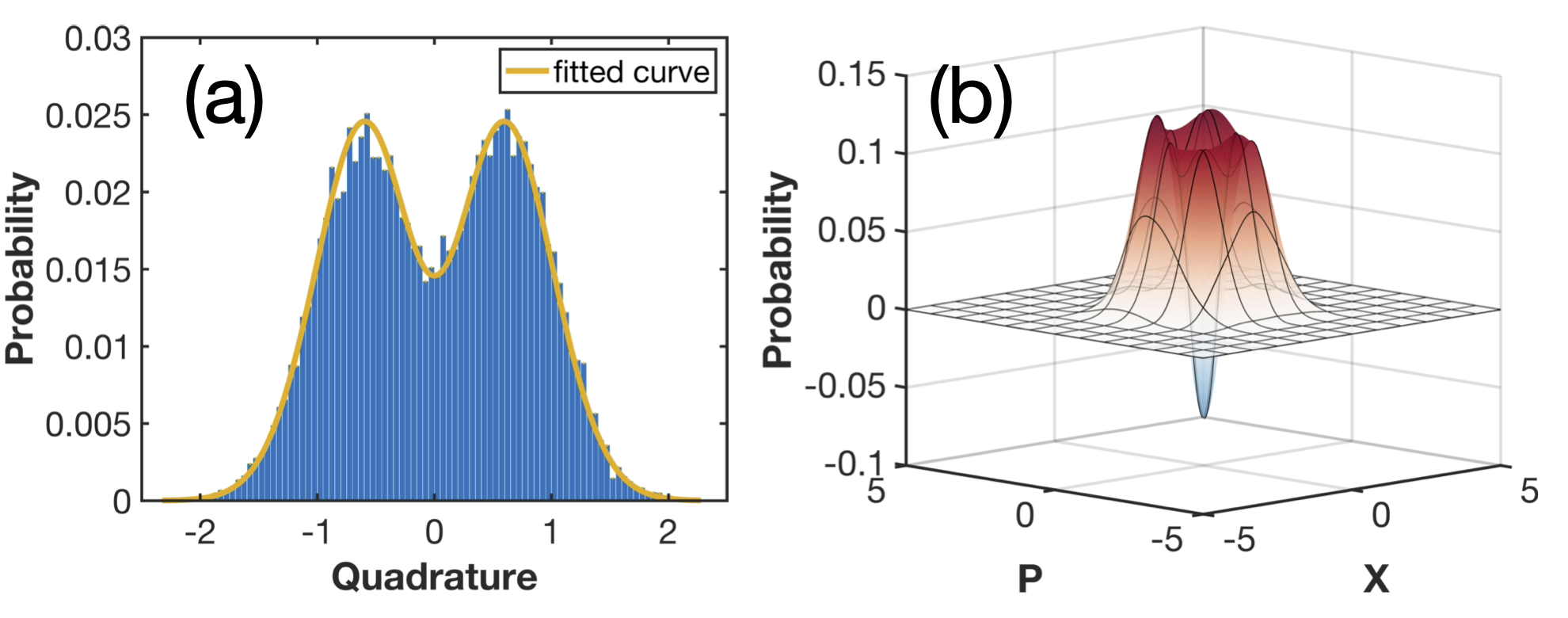}
\caption{(a) The measured probability density in the quadrature ($X$) from homodyne tomography data, with SPDC pump power at 3 mW. (b) The corresponding Wigner distribution function in the phase space. Here, the fitting curve for the probability density (in Yellow color) is fitted by Eq.~(\ref{eqfit}) with $\eta = 0.631$.}
\label{Fdata}
\end{figure}
\begin{figure}
\includegraphics[width=8.0cm]{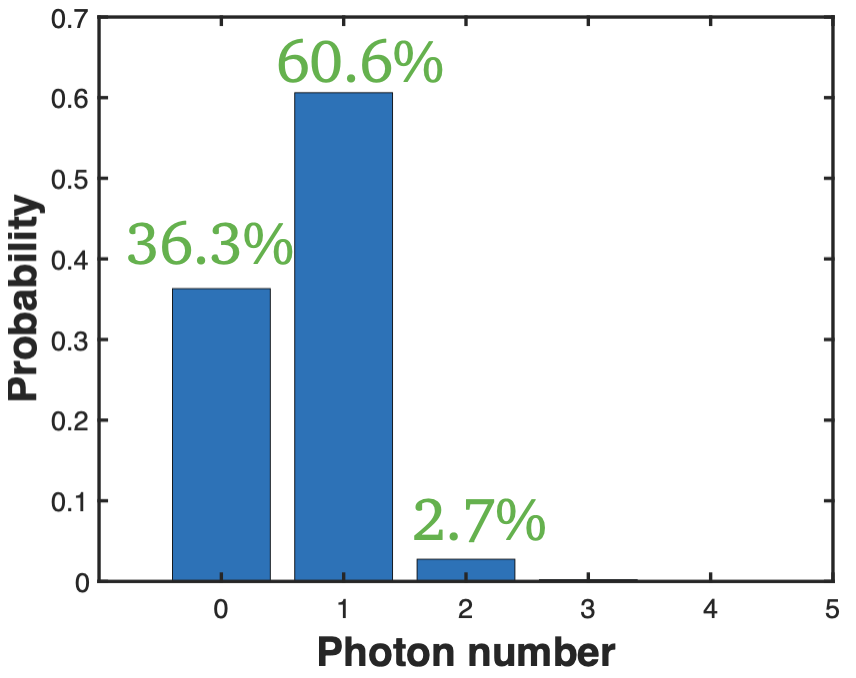}
\caption{The predicted probability distribution generated from our histogram-based QST from the same  measured homodyne tomography data shown in Fig.~\ref{Fdata}(a), with SPDC pump power at 3 mW. Here, we have $w_0 = 0.363$, $w_1 = 0.606$, and $w_2 = 0.027$, corresponding to vacuum, single-photon, and two-photon Fock states, respectively.}
\label{Fprob}
\end{figure}

\section{Results and Discussions}
\label{result}
To verify the validity of our histogram-based QST, first we compare directly with the measured probability density in the quadrature ($X$) from homodyne tomography data.
As shown in Fig.~\ref{Fdata}(a), with the SPDC pump power at 3 mW, the recorded homodyne data illustrates clearly a non-Gaussian probability distribution. By assuming the light field has the form:
\[
|\psi\rangle=(1-\eta)|0\rangle+\eta|1\rangle,
\]
as a noisy single-photon state $|1\rangle$ coupled to the vacuum $|0\rangle$, with the weighting factor $\eta$~\cite{single}.
The corresponding probability distribution function has the form 
\begin{eqnarray}
P(X;\eta)=\sqrt{\frac{2}{\pi}}[1-\eta(1-4X ^2)]e^{-2X^2},
\label{eqfit}
\end{eqnarray}
which gives the best fitting curve depicted in Yellow color, see Fig.~\ref{Fdata}(a), with $\eta = 0.631$. 

With the Wigner-transform, $ {\cal W}[\hat O](x,p) =  \int_{-\infty}^\infty dy\; O(x-\frac{y}{2},x+\frac{y}{2})\; {\rm e}^{\frac{{\rm i}}{\hbar} p y}$ for a single-mode operator given in coordinate representation~$\langle x-y| \hat O | x+y \rangle = O(x-y,x+y)$~\cite{Hancock_EJP04,Cohen_LectureNotes18}, in Fig.~\ref{Fdata}(b) we show the  corresponding Wigner distribution function in the phase space. 
A dip in the origin can be clearly seen, representing the negativity in the Wigner's quasi-probability distribution as a signature of single-photon Fock states.

 In Fig.~\ref{Fprob}, with the same  measured homodyne tomography data shown in Fig.~\ref{Fdata}(a), i.e., the SPDC pump power at 3 mW, we show the predicted probability distribution generated from our histogram-based QST.
Here, in addition to vacuum state $|0\rangle$ and single-photon Fock state $|1\rangle$,  we also take possible two-photon Fock state $|n = 2\rangle$ into  consideration. 
The resulting photon number distribution gives  $w_0 = 0.363$, $w_1 = 0.606$, and $w_2 = 0.027$, corresponding to vacuum, single-photon, and  two-photon Fock states, respectively.
The discrepancy between the direct fitting and our histogram-based QST, see Fig.~\ref{Fdata}(a) and Fig.\ref{Fprob}, comes from the small portion in the two-photon Fock states.

\begin{figure}[t]
\includegraphics[width=8.4cm]{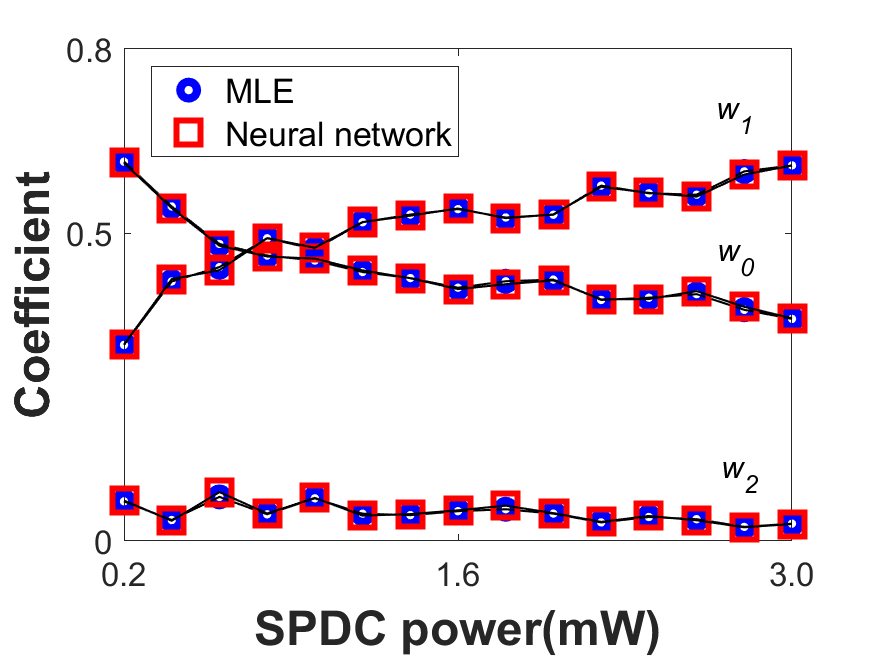}
\caption{Photon number distributions ($w_0$, $w_1$, and $w_2$) versus different SPDC pump power, compared with two different methods:  maximum likelihood estimation (MLE) in `circles' and  our histogram-based neural network in  `squares'.
For the histogram-based neural network, the input bin number is $N = 50$.}
\label{Fcoef}
\end{figure}

As a benchmark, in Fig.~\ref{Fcoef} we also apply MLE method to verify the experimental data at different SPDC pump power.
Here,  both MLE and neural network generate a tiny value for the  three-photon Fock state, i.e.,   $w_{3} < 10^{-13}$, confirming that at most only up to $w_{2}$ (corresponding to two-photon Fock states) is non-negligible. 
With an increment in the SPDC pump power, the coefficient $w_1$ for single-photon Fock states increases; while the coefficient $w_0$ for vacuum states decreases.
As shown in Fig.~\ref{Fcoef}, both two approaches exhibit an almost the same curve, thereby indicating the equivalence and accuracy of these two estimations.
To our surprise, at the same time our SPDC process inside a cavity also produces a small portion of two-photon Fock states, i.e., the average value of $w_2 = 0.044$~\cite{single}.

To avoid the overfitting problem in applying machine learning, we start with the simplest single-layer shallow neural network (only with 50 neurons). 
As we do not apply any complicated structures, the only factor changes the shallow neural network is the input size, which depends on how many discrete points are taken for the quadrature probability density.  For the tests, we have increased the input size to $75$ and $100$, but the resulting  fidelity both generate  $0.999$ without showing significant improvements. 

We want to remark that using a finer discretization also requests  the increment in  the number of input quadrature points. In other words, more data needs to be obtained in the experiment, which  reduces the overall tomographic reconstruction efficiency due to the speed of data collection. In our single-photon experiment, even though we only collected $8,000$ quadrature points,  our current setting can achieve the target, which is also verified with the maximum likelihood estimation.
As a comparison, we  also apply the CNN architecture developed for squeezed states~\cite{PRL-22},  to our single-photon experiments. However, the resulting fidelity can only achieve $0.95$ due to the intrinsic overfitting problem by applying Gaussian states to map non-Gaussian Fock states.

\begin{figure}[t]
\includegraphics[width=8.0cm]{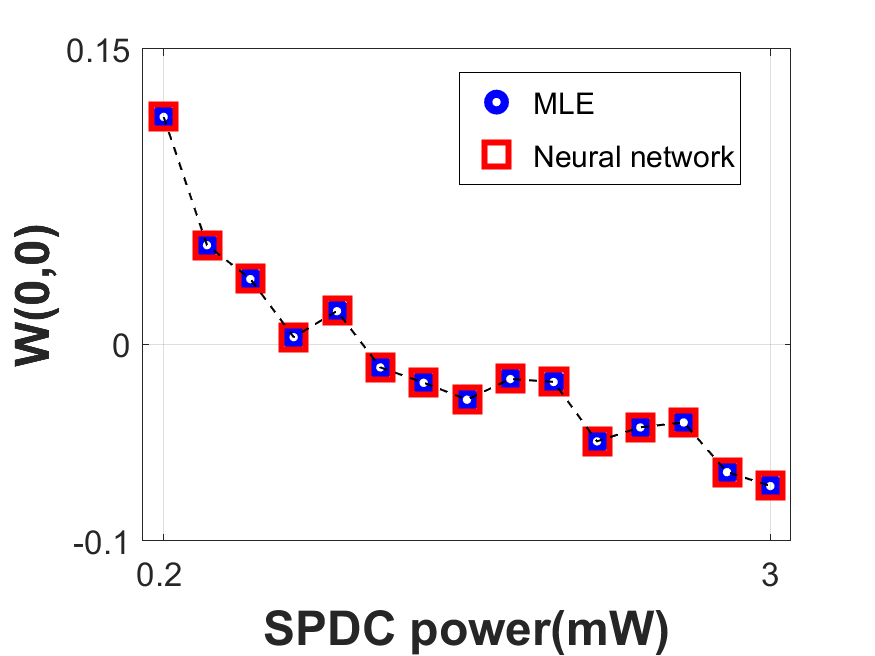}
\caption{Negativity in the Wigner function, $W(0,0)$,  versus different SPDC pump power, compared with two different methods:  maximum likelihood estimation (MLE) in `squares'  and  and  our histogram-based neural network in  `squares'.
For MLE, $W(0,0)$ is calculated after the reconstruction on the  quantum state; while the neural network directly estimates $W(0,0)$ by Eq.~(\ref{eqW00}) form the quadrature histogram.}
\label{Fneg}
\end{figure}

In addition to the photon number probability estimation, our histogram-based neural network can also predict directly the target parameters, without the reconstruction on the full quantum state.
Here, we focus on the negativity in the Wigner's quasi-probability distribution, which manifests the most non-classical signature of single-photon Fock states.
Analytically, the value in the origin of  Wigner function  $W(0, 0)$ has the form 
\begin{eqnarray}
W(0,0)=\frac{1}{\pi}\sum_{n=0}^{2}(-1)^{n}w_{n}L_{n}(0),
\label{eqW00}
\end{eqnarray}
 with the Laguerre polynomial $L_n (x)$~\cite{9}.
In Fig.~\ref{Fneg}, we illustrate the powerful feature in our histogram-based neural network by directly infer the negativity in the Wigner function, $W(0,0)$,  versus different SPDC pump power.
Here, we also compare the results generated from  two different methods:  MLE-QST and our histogram-based neural network.
It is noted that in MLE, $W(0,0)$ is calculated after the reconstruction on the  quantum state.
Nevertheless, our  neural network directly estimates $W(0,0)$ directly by using Eq.~(\ref{eqW00}) form the quadrature histogram.
As one can see, again, our ML parameter estimation gives almost the same results as that from MLE.

By considering SPDC experiments in the low-intensity limit, the condition to have a negative value in $W(0,0)$ happens at $w_{1}=0.5$, corresponding to our SPDC pump power at  0.8 mW.
As shown in  Fig.~\ref{Fneg}, our histogram-based neural network, also confirmed by MLE, precisely estimate the negativity happens when the SPDC pump power exceeds 0.8 mW.

Last but not least, due to the perfect agreement between the results from MLE method and our  histogram-based QST,  we choose  50 bins as a good estimation. 
Unlike MLE method relying on the iteration algorithm, we can have a reusable prediction map from our neural network. 
This lightweight feature  makes it easier to install such an inference function on edge devices like FPGA. 
Most of time-consuming task in our approach  is the pre-processing, i.e., the histogram binning process, which takes bout 0.01 seconds. 
However, it only takes about 3 msec to subsequently go through such a tiny $50 \times 3$ network for inference. 
The total time consumed is about 0.01 + 0.003 seconds.

\section{Conclusion}
\label{conclusion}
In summary, we develop a neural network enhanced single-photon Fock state tomography and apply it to the heralding single-photon source from spontaneous parametric down conversion (SPDC) process experimentally.
Instead of tackling on the raw quadrature data, which needs a lot of computational cost but infers a limited fidelity up to $0.96$, our histogram-based quantum state tomography (QST) keeps the fidelity as high as $0.999$.
Moreover, target parameters, such as the photon number distribution and the negativity in Wigner function, can be directly predicted, without dealing with the density matrix in a higher dimensional Hilbert space. 

Through the validation with the experimentally measured data acquired from the balanced homodyne detectors, perfect agreement to the results obtained by maximum likelihood estimation (MLE) is also clearly demonstrated. 
Our machine-learning enhanced QST can be easily installed on edge devices such as FPGA as an in-line diagnostic toolbox for all the possible applications with single photons.
Moreover, this fast and easy-to-install methodology helps us with a better understanding on quantum optics experiments with non-Gaussian states,   such as two-photon Fock state tomography~\cite{two-photon}, photon-added squeezed states~\cite{cat}, tomographic test of Bell’s inequality~\cite{Bell}, and 
the reconstruction of non-classicality~\cite{decoherence}.

\section*{Acknowledgements}
This work is partially supported by the Ministry of Science and Technology of Taiwan (Nos
112-2123-M-007-001, 112-2119-M-008-007, 112-2119-M-007-006), Office of Naval Research Global, the
International Technology Center Indo-Pacific (ITC IPAC) and Army Research Office, under Contract
No. FA5209-21-P-0158, and the collaborative research program of the Institute for Cosmic Ray
Research (ICRR) at the University of Tokyo.

\end{document}